\documentclass[preprint2]{aastex}

\slugcomment{vs. 9: 2009 Mar 26: replace figures with revised uncertainties for fpy}

\def\ltsima{$\; \buildrel < \over \sim \;$}
\def\simlt{\lower.5ex\hbox{\ltsima}} 
\def\gtsima{$\; \buildrel > \over \sim \;$}
\def\simgt{\lower.5ex\hbox{\gtsima}} 
\def\arcsec{\hbox{$^{\prime\prime}$}}

\def\HST{{\it HST}}
\def\VLA{{\it VLA}}
\def\VLBA{{\it VLBA}}
\def\Chandra{{\it Chandra}}
\def\fpyp{$fpy^{+}$}
\def\fpym{$fpy^{-}$}
\def\fpy{{\it fpy}}

\shorttitle{Variability in the M87 Jet}
\shortauthors{Harris et al.}

\begin{document}


\title{Variability Timescales in the M87 Jet: Signatures of $E^2$
Losses, Discovery of a Quasi-period in HST-1, \& the Site of TeV
Flaring}



\author{D. E. Harris} 
\affil{Smithsonian Astrophysical Observatory, 60
Garden St., Cambridge, MA 02138} \email{harris@cfa.harvard.edu}

\author{C. C. Cheung} 
\affil{NASA Goddard Space Flight Center,
Astrophysics Science Division, Greenbelt, MD 20771} 

\author{{\L}ukasz Stawarz\altaffilmark{1}}
\affil{Kavli Institute for Particle
Astrophysics and Cosmology, Stanford University, Stanford, CA 94305}

\author{J. A. Biretta}
\affil{Space Telescope Science Institute, 3700 San Martin Drive, Baltimore, MD 21218}

\author{E. S. Perlman} \affil{Physics and Space Sciences Department,
Florida Institute of Technology, 150 West University Boulevard,
Melbourne, FL 32901}

\altaffiltext{1}{Astronomical Observatory, Jagiellonian University, ul. 
Orla 171, 30-244 Krak\'ow, Poland.}



\begin{abstract}

We investigate the variability timescales in the jet of M87 with two
goals.  The first is to use the rise times and decay times in the
radio, ultraviolet and X-ray lightcurves of HST-1 to constrain the
source size and the energy loss mechanisms affecting the relativistic
electron distributions.  HST-1 is the first jet knot clearly resolved
from the nuclear emission by \Chandra\ and is the site of the huge
flare of 2005.  We find clear evidence for a frequency-dependent
decrease in the synchrotron flux being consistent with $E^2$ energy
losses.  Assuming that this behavior is predominantly caused by
synchrotron cooling, we estimate a value of 0.6 mG for the average
magnetic field strength of the HST-1 emission region, a value
consistent with previous estimates of the equipartition field.  In the
process of analyzing the first derivative of the X-ray light curve of
HST-1, we discovered a quasi-periodic oscillation which was most
obvious in 2003 and 2004 prior to the major flare in 2005.  The four
cycles observed have a period of order 6 months.  The second goal is
to search for evidence of differences between the X-ray variability
timescales of HST-1 and the unresolved nuclear region (diameter
$<~0.6^{\prime\prime}$).  These features, separated by more than 60
pc, are the two chief contenders for the origin of the TeV variable
emissions observed by HESS in 2005 and by MAGIC and VERITAS in 2008.
The X-ray variability of the nucleus appears to be at least twice as
rapid as that of the HST-1 knot.  However, the shortest nuclear
variability timescale we can measure from the {\it Chandra} data
($\leq 20$\,days) is still significantly longer than the shortest TeV
variability of M~87 reported by the HESS and MAGIC telescopes ($
1-2$\,days).

\end{abstract}


\section{Introduction} 

We have been monitoring the jet of M87 in the X-rays with \Chandra\ and 
the ultraviolet (UV, at $\lambda$=220 nm) with \HST\ since 2002 
Jan, and in the radio since 2003 (\VLA, primarily at 15 GHz) and 2005 
(\VLBA, primarily at 1.7 GHz). Previous papers from this project include 
Paper I reporting our first results (Harris et al. 2003), Paper II which 
focused on the \HST\ data (Perlman et al. 2003), Paper III which was 
mainly on the X-ray lightcurve of HST-1 which delineated the massive flare 
in 2005 (Harris et al. 2006) and Paper IV, the \VLBA\ results showing 
superluminal proper motions in HST-1 (Cheung, Harris \& Stawarz 2007).

In this paper (V of the series), we present an analysis of the
lightcurves for the nucleus, the jet knot HST-1 which is 0.86\arcsec\
from the core and which is the site of the massive X-ray, UV, and
radio flare described in Paper III, knot D, and knot A. Knot A was
mainly used as a control source since we do not expect short timescale
variability because it is well resolved in all bands. The lightcurve
of Knot D illuminates the effects of HST-1 on adjacent regions since
it appears to have very little if any intrinsic variability on the
timescales of interest here.  In \S~\ref{sec:xray}, \ref{sec:uv}, and
\ref{sec:radio} we describe the data and the analyses methods.  In
\S~\ref{sec:rise} we use the rising segments of the lightcurves to
derive upper limits on the size of the emitting regions, and in
\S~\ref{sec:decay} we examine the decay timescales of HST-1 in X-rays,
UV, and radio bands in order to isolate signatures of $E^2$ losses.
We describe newly discovered oscillations in the brightening and
fading of HST-1 in \S~\ref{sec:impulse}.  Finally, in
\S~\ref{sec:compare}, we discuss the evidence for short timescales in
the X-ray variability of the nucleus and HST-1 which is relevant to
the question of the location of the variable TeV emission reported by
the HESS collaboration \citep{ahar06}, MAGIC \citep{albe08}, and
VERITAS \citep[e.g.][]{veritas08}.  Some preliminary results from this
work were reported in \citet{harr08}.

We take the distance to M87 to be 16 Mpc \citep{ton91} so that 
1$^{\prime\prime}$ corresponds to 77 pc.  Throughout this paper we 
assume that the radio to X-ray emission from all parts of the M87 jet 
comes from synchrotron emission, as argued in our previous papers (I \& 
III).  In particular, we assume the X-ray nuclear emission is dominated by
synchrotron emission from components of the inner (unresolved) jet rather
than by thermal processes associated with the accretion disk around the
central black hole.

\section{The \Chandra\ X-ray Data}\label{sec:xray} 

Since 2002, we have observed M87 with \Chandra\ 6--7 times each observing 
season with 5 ks exposures typically separated by six weeks. Additionally 
in 2005, near the maximum of the X-ray lightcurve of HST-1, we scheduled 
weekly observations to constrain shorter timescale variability 
(epochs Ya to Yg; see Table~\ref{tab:obs}).  After the report of 
variable TeV emission in the 2005 HESS observations of M87 by 
\citet{ahar06}, we obtained Director's Discretionary Time (DDT) 
observations to sample the X-ray lightcurves on $\sim$2--3 day timescales 
during two 'dark time' fortnights in Feb. and Mar. of 2007 when TeV 
observations were scheduled (epochs Ys to Zb). A total of 61 observations 
have been obtained from these programs thus far.


For details of our reduction procedures, see Papers~I \& III.
Briefly, we use a 1/8th segment of the back illuminated S3 chip of the
ACIS detector aboard \Chandra.  This permits us to have a frame time
of 0.4s with 90\% efficiency.  Although this setup was essentially
free of pileup when \citet{wils02} tested various options during 2000
July, with the advent of the ever increasing brightness of HST-1,
pileup \citep{davis01} became a major problem so we switched to a
detector based measure of intensity: keV/s.  This approach uses the
event 1 file with no grade filtering (so as to recover all events
affected by 'grade migration') and we integrated the energy from 0.2
to 17 keV so as to recover all the energy of the piled events.  Other
uncertainties for piled events comes from the on-board filtering, the
'eat-thy-neighbor' effect, and second order effects such as release
of trapped charge (see \S~\ref{sec:pileupprob}).

\subsection{Photometry}

\begin{figure} 
\plotfiddle{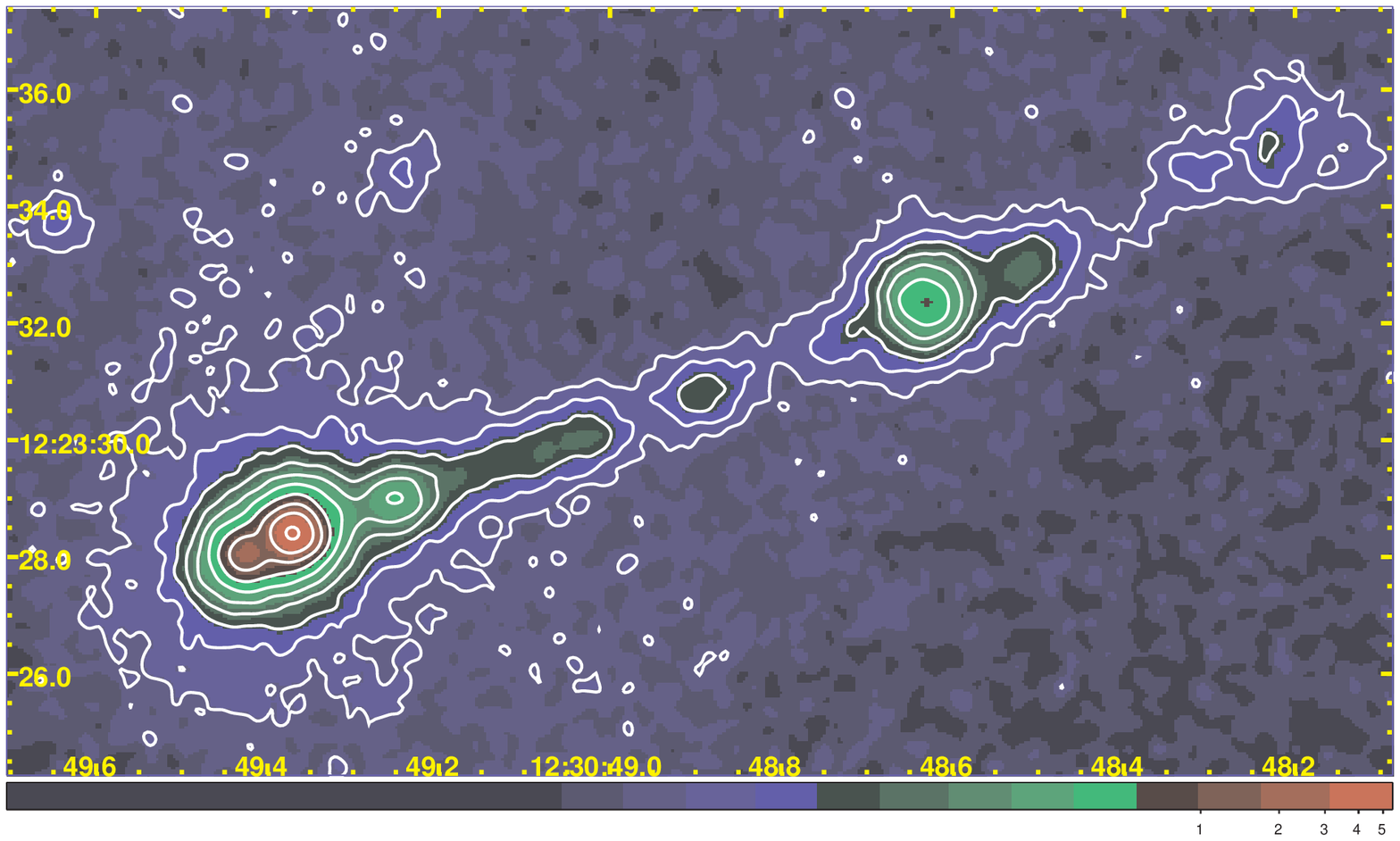}{0cm}{0}{210}{128}{0}{2200}
\plotfiddle{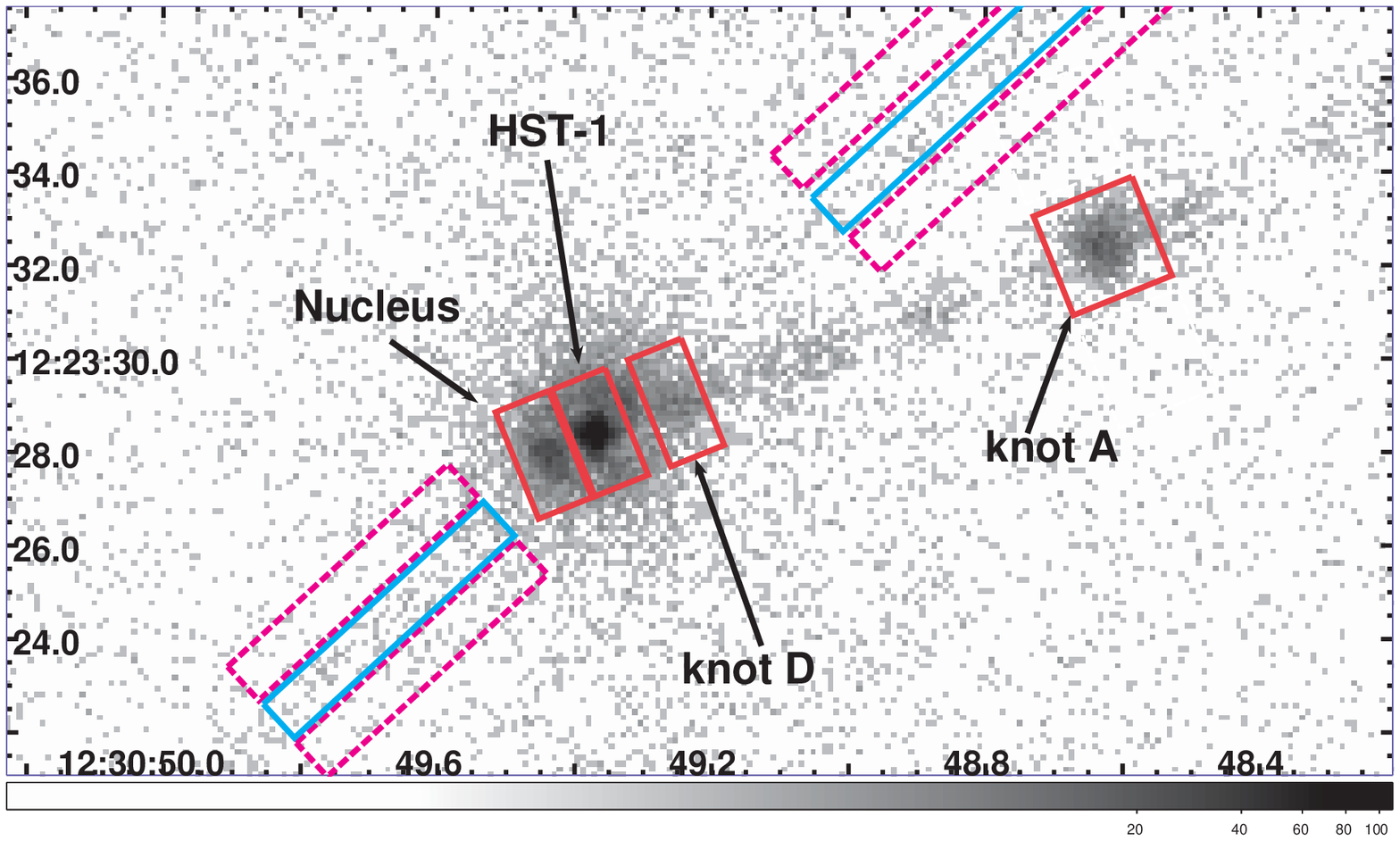}{0cm}{0}{210}{150}{0}{-15}

\caption{A \Chandra\ X-ray image constructed from the sum of our data from 2002 to
  2004.  This is a total power image for which each event has been
  multiplied by its energy.  In the top panel, a Gaussian smoothing function of FWHM=
  0.25$^{\prime\prime}$ has been applied.  Contours start at 0.01 eV s$^{-1}$
  per 0.049$^{\prime\prime}$ pixel and increase by factors of two.
  The second panel shows the regions used for photometry.  The four
  rectangular regions in red (from left to right) are the nucleus, the
  flaring knot HST-1 (0.86$^{\prime\prime}$ from the nucleus), knot D, and knot A.  The long thin cyan
  rectangles and the dotted magenta rectangles are used for 'readout
  streak photometry' ('on' and 'background', respectively) -- see \S~\ref{sec:readout}.  The image
  is a 5 ks exposure from 2005 May when HST-1 was close to its peak
  intensity. Pixel randomization has been removed and in this figure panel,
  the events are binned into 0.123$^{\prime\prime}$ pixels.  The
  axes are J2000 coordinates.}\label{fig:regions}
\end{figure}

Although we used small circular apertures for fluxmap photometry in 
Paper I, the basic analysis for this paper adopts the rectangular 
regions used in Paper III so as to encompass more of the point spread 
function (PSF).  The 4 regions of interest (the core, HST-1, knots D and 
A) are shown in fig.~\ref{fig:regions}; we did not use background 
subtraction because the photometric apertures were small.

All events within each rectangle are weighted by their energy and the sum
of these energies, when divided by the exposure times, gives the final
keV/s value used in the lightcurve.  Uncertainties are strictly
statistical, based on the number of counts measured: $\sqrt{N}/N$ and
typically range from 1\% to 5\%.

To analyze and compare timescales, we measure the slope between adjacent 
measurements (and also between every other observation) by calculating the 
ratio, $(I_{\rm 2}-I_{\rm 1}) / \Delta t$, where $I_{\rm 1}$ and $I_{\rm 
2}$ are the intensities at the times $t_{\rm 1}$ and $t_{\rm 2}$ and 
$\Delta t {\rm [yrs]} = t_{\rm 2}-t_{\rm 1}$.  To convert this to a fractional 
change we divide by $\min(I_{\rm 1},I_{\rm 2})$, so the definition of \fpy\ is:

\begin{equation}
fpy = \frac{D}{I_{\rm i}\Delta t},  
\end{equation}

\noindent where $D=(I_{\rm 2}-I_{\rm 1})$.
When the intensity is {\it increasing} ($I_{\rm 1}<I_{\rm 2}$), the
fractional change per year has i=1 and is denoted (when required), as
\fpyp. When the intensity is {\it dropping} ($I_{\rm 1}>I_{\rm 2}$), i=2 and 
we specify by using \fpym.

\noindent (Hereafter, the superscripts are suppressed where the signs are 
implicit). According to this definition, the doubling time for a given 
value of $fpy$ is simply $t_{\rm double}~{\rm [yr]} = 1 / fpy$, thus when 
$fpy = \pm$1, there was a rate of change which would produce a factor of 
two increase or decrease in $I$ in one year.

The uncertainties in each value of $I$ are propagated to the first
derivative by calculating the square root of the sum of the squares of
the errors on intensity.  Denoting $\sigma_{\rm i}$ as the uncertainty of
$I_{\rm i}$, we express the error of \fpy\ as:

\begin{equation}
\sigma(fpy) = \frac{1}{\Delta t}\times\frac{I_{\rm j}}{I_{\rm i}}\times~\sqrt{(\frac{\sigma_{\rm 1}}{I_{\rm 1}})^2 + (\frac{\sigma_{\rm 2}}{I_{\rm 2}})^2}
\end{equation}

\noindent Here i=1, j=2 for the error on \fpyp, and i=2, j=1 for \fpym.

\begin{figure} 
\plotfiddle{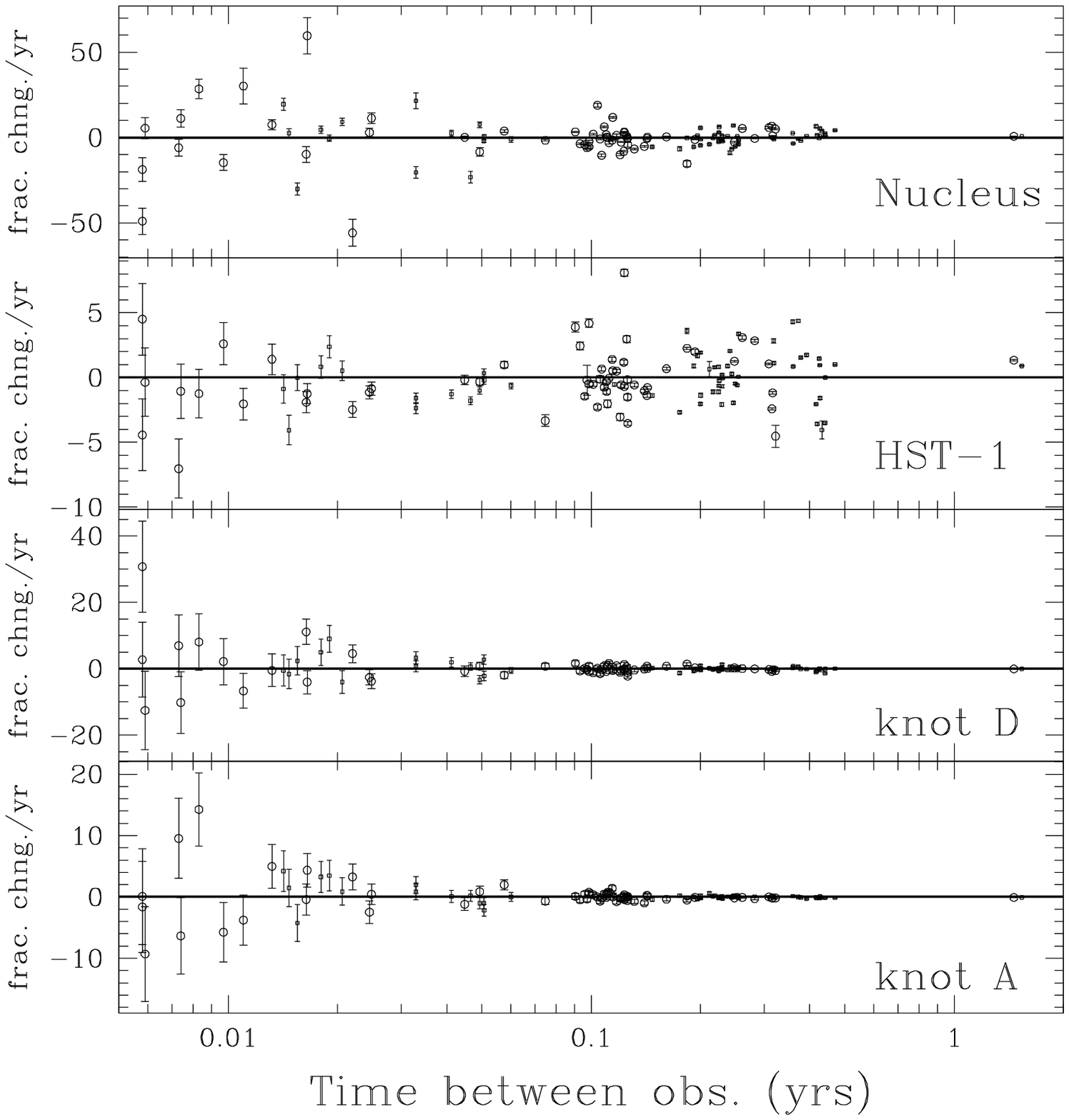}{0cm}{0}{230}{230}{0}{-432}%
\plotfiddle{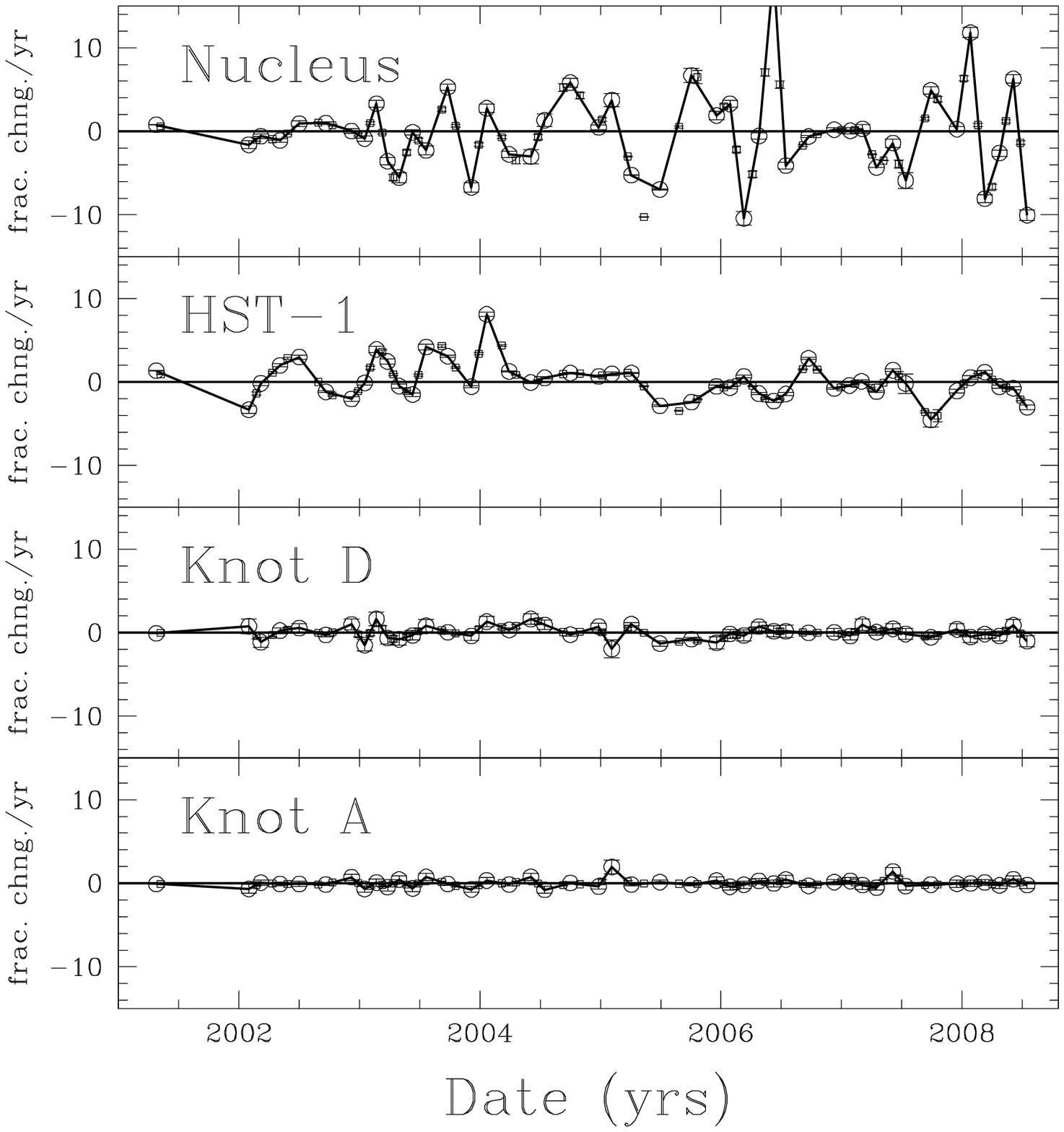}{0cm}{0}{230}{230}{0}{-2200}


\caption{The first derivative of the lightcurves for the nucleus, HST-1, 
knots D and A (top to bottom).  The smaller squares are values from 
intensity pairs which skip the intervening observation and are thus not 
independent of the larger circles (adjacent observations). In the upper 
panel the x axis is the time between observations; in the lower panel it 
is the date. The y axis is fractional change per year. Note the change 
in y scale for different plots in the upper panel, but in the lower 
panel we have set the max/min values to $\pm$~15.  In the lower panel, 
intervals with closely spaced observations have been averaged to avoid 
large error bars arising from small $\Delta$t.}\label{fig:4fpy} 
\end{figure}

The $fpy$ values are plotted in fig.~\ref{fig:4fpy} for the nucleus,
HST-1, knots D and A.  The latter two features serve as controls, and
display the expected behavior for a steady source: at long time
intervals, $fpy$ values are consistent with zero and have small
uncertainties.  The errors increase as $\Delta t$ decreases.  Since
all errors are 1$\sigma$ ($\sqrt{N}/N$), some values for knots D and A
appear to be different from zero, as expected.

In the next section we discuss using the readout streak for estimating
source intensity and in the Appendix (\S~\ref{sec:pileupprob}) we describe
some of the other problems engendered by pileup.

\subsection{Photometry of the Read-out Streak\label{sec:readout}}

The only check we have been able to devise is 'read-out streak
photometry' \citep[see also][]{mar05}.  For this procedure, we isolate
the segments of the readout streak which are not close to the jet.
With long thin rectangles (n$\times$2 pixels; fig.~\ref{fig:regions}),
and adjacent rectangles to measure the background, we estimate the
effective exposure time for the net counts as (\#frames)
$\times~n~\times 41\mu$s, where $n$=length of rectangles in pixels.
In a typical 5 ks observation, we are thus able to get the equivalent
of 30 to 40 seconds worth of continuous clocking mode data which are
free of pileup effects but suffer from poor s/n because of the
necessarily large background area.  While the general behavior of the
streak photometry is consistent with the expectation that it would
become increasingly larger than the standard photometry as the
intensity of HST-1 increased (more pileup and more on-board
rejection), there are a few unexpected departures from this
behavior. In particular, there are some high values where both methods
give the same intensity, and there is one segment of low intensity
where the streak photometry is significantly less than the standard
photometry.  For these reasons we rely on our 'standard' photometry
and relegate the results of the readout streak photometry to the
status of 'caveats' and 'alternate possibilities'.

\section{The UV Data}\label{sec:uv} 
%

The UV data used in this paper were obtained from a series of \HST\
proposals (Biretta, PI) which were synchronized with the \Chandra\
observations although for various reasons not every \Chandra\
observations has a corresponding \HST\ observation.  The data were
reduced with the usual procedures and will be described in a separate
paper (Biretta, in preparation).

\section{The Radio Data}\label{sec:radio} 

The \VLA\footnote{The National Radio Astronomy Observatory is a facility
of the National Science Foundation operated under cooperative agreement
by Associated Universities, Inc.} 15 GHz observations were obtained as
part of our multi-frequency program coordinated with the \Chandra\ and \HST\
monitoring and began mid-2003. In each \VLA\ cycle, we observed M87 in
three 8 hr runs (program codes AH822, AH862, AH885, AC843): two in A
array followed by one in B array. The beam sizes were typically around
0.15$"$ and 0.4$"$, respectively. The longer gaps occur during C and D
array configurations when the angular resolution is not sufficient to
separate the nucleus and HST-1.

The observations utilized two adjacent 50 MHz wide intermediate 
frequencies centered at 14.94 GHz. A total of 1--1.5 hrs of on-source 
time was split between 9-10 scans over each 8 hr run to obtain good 
$(u,v)$ coverage. The data were calibrated in AIPS \citep{bri94} with 
the flux density scale set using scans of 3C286 and the initial phase 
corrections determined with a nearby calibrator. Subsequent phase and 
amplitude self-calibration was performed using the Caltech DIFMAP 
package \citep{she94}.

An additional archival B-array dataset (BK073) was analyzed to give us
the data point in January 2000, before our monitoring began. The
observation used an identical setup to ours but with only four 3.5 min
scans obtained. This was sufficient to detect a faint (2.8 mJy) feature
at the position of HST-1, thus providing a baseline measurement to our
subsequent ones.

\section{Estimating the Size of Emitting Volumes from Rise Times}\label{sec:rise}

As described in earlier papers of this series, the rise of the light 
curve can give estimates of the source size so long as the beaming 
parameters -- the bulk Lorentz factor of the jet ($\Gamma$), the angle 
between the jet and our line of sight ($\theta$), and the Doppler 
beaming factor ($\delta$) -- are not changing.

For the standard analysis of the X-ray light curve of HST-1, the
largest observed slope occurred early in 2004 and had a value of
$fpy=8$ (fig.~\ref{fig:4fpy}); the intensity doubled during our 6-week
sampling interval.  Thus we are able to set an upper limit to the
characteristic size of the emitting region in the jet frame of
diameter $\leq 0.12 \delta$ light years (45 light days).
Significantly larger values of $fpy$ occur for the HST-1 light curve
generated from the readout streak photometry (\S~\ref{sec:readout}),
but these have large uncertainties so do not yield useful limits on
the source size.

The smallest directly measured value for the size of HST-1 comes from the 
\VLBA\ beamsize of 3 mas which corresponds to 0.7 light years (Paper IV).  
If the X-ray and radio emitting volumes were to be one and the same (i.e., 
the upstream end of HST-1), and if the radio size is actually similar to 
our \VLBA\ resolution, then $\delta$ would be of order 5, a value similar 
to that estimated in our previous papers.

\begin{figure} 
\includegraphics[angle=0,scale=.40]{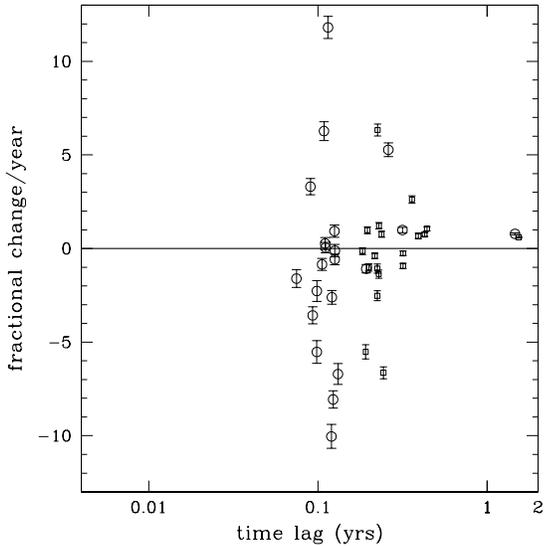}
\includegraphics[angle=0,scale=.40]{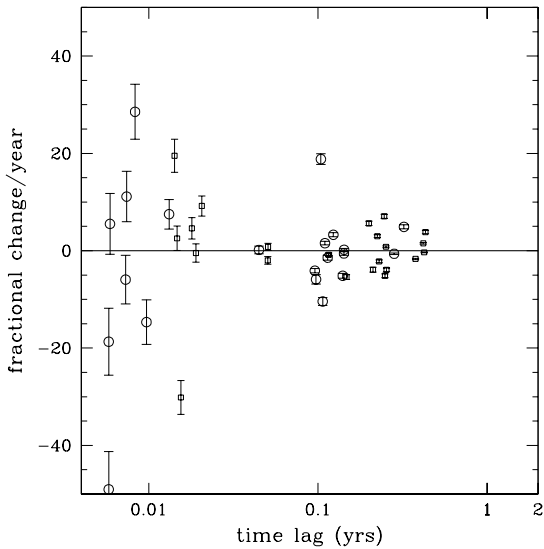}
\caption{\fpy\ values for the nucleus.  The larger open circles come
from adjacent observations whilst the smaller squares are data pairs
formed by skipping the adjacent observation.  First panel: the period
from 2000-2004 and 2008 when the HST-I intensity was low ($<$2 keV/s).
Second panel: the period 2006-2007 when the intensity of HST-1 was
between 2 and 4 keV/s. \label{fig:cchlowmed}}
\end{figure}

Because of the second order effects which were prevalent when HST-1
was strong, we have restricted our analyses of the nuclear $fpy$ data
to (a) 2000-2003 plus 2008 when HST-1 was at a low intensity (less
than 2 keV/s), and (b) 2006-2007 when the HST-1 intensity was at an
intermediate value (2 to 4 keV/s).  The data between 2 and 4 keV/s are
used because this time interval includes the closely spaced
observations of 2007.  We believe the second order effects for this
intensity regime are minimal because although the light curve of the
nucleus shows a very obvious featue corresponding to the peak of HST-1
in 2005, there is no apparent correlation between the intensity of the
nucleus and that of HST-1 for the 2006-2007 data.  These data are
shown in fig.~\ref{fig:cchlowmed}.

During low intensity intervals, the maximum value of \fpyp\ was about
12 (a light travel time of $30 \delta$ light days), although there
were no closely spaced observations during these periods so that larger
\fpyp\ values may have been missed because of inadequate sampling.
When HST-1 was at intermediate levels ($\sim$2--4 keV/s), we had the
closely spaced observations during 2007 Feb and Mar from the DDT
project and the maximum \fpyp\ for the core was 28.5$\pm$5.6, but we
take a value of 19 (light travel time: $19\delta$ light days) as a
characteristic value since the second largest value was 18.8$\pm$1.1.

\section{Analysis of the Decay Phases of the Lightcurves of 
HST-1}\label{sec:decay}

\begin{figure}[tb] 
\includegraphics[angle=0,scale=.40]{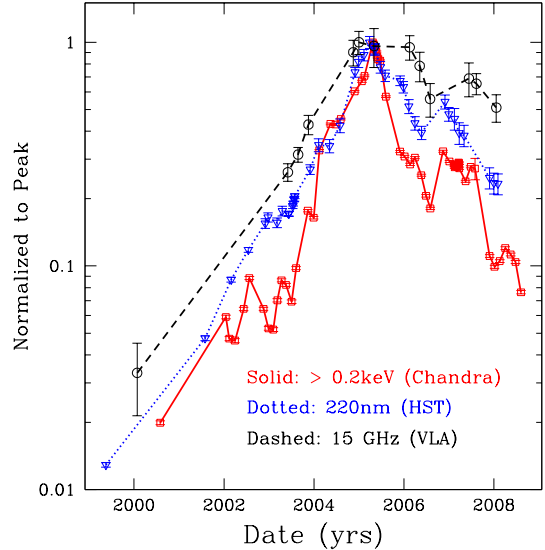}
\caption{The X-ray, UV, and radio lightcurves of HST-1.  The intensity
is plotted on a log scale to demonstrate the overall conformity
between bands.  Each curve has been normalized by setting the peak
value to unity so as to permit visual comparison of the decay.  Peak
values are 12.417 keV/s (X-ray); 0.596 mJy (UV); and 0.084 Jy
(15GHz).}\label{fig:roxlc}
\end{figure}

In this section, we analyze the decay of the light curves of HST-1 at
X-ray, UV, and radio wavelengths (fig.~\ref{fig:roxlc}).  We will not
attempt to make a parallel investigation of the nuclear emission
because we have no information as to the size of the emitting volume
or its geometry, the interpretation of the UV data would be
problematic, and although likely, it remains to be demonstrated that
the nuclear X-ray emission is actually non-thermal emission from the
inner jet rather than from some thermal process associated with the
accretion disk or its environs.

The decay of the light curve may be caused by several effects but has the 
potential to reveal which processes are dominant.  If all bands have 
similar rates of decreasing intensity, the most likely cause is either a 
change in the beaming factor (as might arise from a change in $\theta$) or 
a general expansion which reduces the energy of all electrons according to 
their energy (the so called ``$E^1$ losses'') as well as reducing the 
magnetic field strength.  Note however that if the 
emitted spectrum is not a simple power law, but for example steepens at 
high frequencies, then a simple expansion may produce a much stronger 
decay at high frequencies both because the previously 'viewed' electrons 
for a fixed observing band are now replaced by the fewer (previously 
higher energy) electrons and also because of the weaker B field, the fixed 
observing band now comes from an even higher energy segment of the 
electron distribution.

Our preliminary analysis of Paper III indicated that the initial
decrease of the major flare had a similar timescale as the preceding
rise, and that this might be indicative of either a changing $\delta$,
or a compression and subsequent expansion.  However, it has become
clear that although the UV and X-ray light curves appeared to have a
similar behavior initially, there are instrumental effects present
which were not recognized, there are significant differences in the UV
and X-ray decays and the radio intensity did not conform to the rapid
decay seen at higher frequencies (fig.~\ref{fig:roxlc}).

We investigate two aspects of this problem.  First, we examine the
behavior of the UV and radio lightcurves at times when large rates of
decay are observed at X-rays, and second we compare the $fpy$ values
between bands without regard to when they occurred.  Both of these
approaches suffer from the sparser sampling in the radio.  Although
most of the UV data were obtained within a week of the \Chandra\
observations, not every \Chandra\ observation has a corresponding \HST\
observation.

\subsection{Comparison of particular time segments}

This approach is based on the assumption that all 3 wavelength bands
come from the same emitting volume.  If however, the emitting volume is
'layered' (e.g., concentric spheres with longer wavelengths coming from
larger volumes), then it would be possible to have different
characteristic decay times for each layer (assuming different values
of the magnetic field strength), and the decays in the light curves
would not have to happen at the same time since the cessation of injection
of particles and fields would not necessarily be simultaneous in all emitting
volumes.

The most significant (i.e., without excessively large uncertainties) 
decays in the X-ray lightcurve occurred in (a) 2002.0 ($fpy = -3.32 \pm 
0.45$), (b) 2005.5 ($fpy = -3.51 \pm 0.08$), and (c) 2007.7 ($fpy = -4.53 \pm 
0.85$).

For event (a), we have no relevant radio values of $fpy$ and the UV
sampling was not sufficient: the UV intensity rose from 28 to 52 $\mu$Jy
between 2001.58 and 2002.16.  The decay in the X-rays occurred between
2002.044 and 2002.119.

Event (b) occurred after the peak of the giant flare.  This is perhaps
the best example of what might be expected from $E^2$ losses affecting
the electron energy distribution (see fig.~\ref{fig:roxlc}).  However,
as mentioned above, we can't rule out the possibility of an expansion
of the source if the X-rays are coming from a segment of the electron
distribution that is falling more rapidly than the power law
connecting the UV to the X-ray.  The $fpy$ value for the UV is
--0.85$\pm$0.86.

The third notable decline in the X-ray light curve came during the 3 
months in 2007 whilst M87 was behind the Sun (Aug. - Nov.).  
Unfortunately, the UV monitoring had a much longer gap (almost 7 months) 
so the relevant data for a direct comparison do not exist.

\begin{figure}[tb] 
\includegraphics[angle=0,scale=.40]{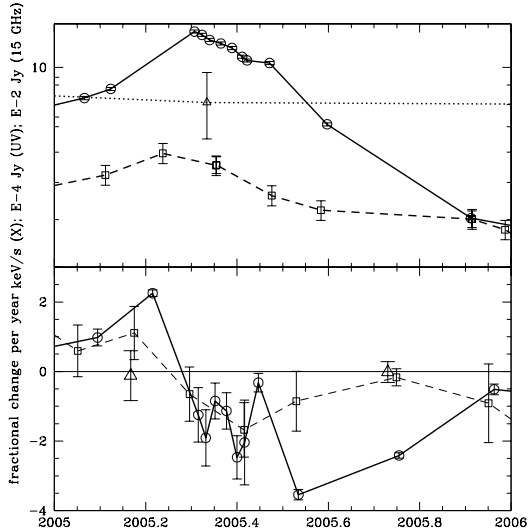}
\caption{Multiband data for HST-1 during 2005.  The top panel shows
the lightcurves: circles for X-ray; squares for UV with dashed line;
and triangles for the 15 GHz data (dotted line).  The lower panel shows
the corresponding $fpy$ values, the first derivative of the light
curves.  Coding is the same as the upper panel except no line is added
to the 2 radio data, both of which are consistent with zero.  Note the
shoulder in the UV lightcurve following the peak.  This can also be
seen in fig.~\ref{fig:roxlc}, and leads to the much smaller values of
dI/dt for the UV than for the X-ray.  The radio lightcurve is
essentially 'flat topped' and shows no change during the
year.}\label{fig:rox2005}
\end{figure}

If the X-ray \fpym\ value of --3.5 during the main decay phase ('event
b', fig.~\ref{fig:rox2005}) was dominated by $E^2$ losses, we can make
an order of magnitude estimate for the magnetic field strength.  Note
that we do not consider here a detailed evolution of the synchrotron
flux produced by the radiatively cooling electron energy distribution
as considered in, e.g., \citet{kard62}, but we can approximate the
cooling time required to drop the intensity by a factor of two by
assuming that this decay is caused by a factor of two fewer electrons
with energies providing the bulk of the observed X-rays between 0.2
and 6 keV.  Further, we assume the exponent of the electron
distribution, p=2$\alpha_x$+1=3.4, which is our best estimate for the
spectrum of HST-1 in the X-ray band \citep{pap3}.  Since dE/dt to
first approximation shifts the power law distribution, N(E) to lower
energies, we require an energy shift of 2$^{-1/3.4}$ or 0.8.  We then
ask what magnetic field strength, B is required to produce a 20\%
energy loss in the time, $\tau$, it takes for the intensity to fall by
a factor of two.  The observed time, $\tau_o$ is 1/3.5=0.28yr; in the
jet frame $\tau'$=$\delta\tau_o$ where $\delta$ is the beaming factor.

With the standard equations (3.28 and 3.32) from \citet{pac70},
setting $\tau'$dE/dt=0.2E (where E is the energy of the electron),
and changing jet frame parameters to the observer frame, we find

\begin{equation}
B \approx 0.5\delta^{-1/3} \, \left({\tau_o \over {\rm yr}}\right)^{-2/3}
\, \left({\varepsilon \over {\rm keV}}\right)^{-1/3} \, {\rm mG} \,.
\end{equation}

\noindent
Here $\tau_o/{\rm yr} \equiv 1 / fpy^-$ is the observed time for the
intensity to drop by a factor of two and $\varepsilon$ is the
characteristic energy of the X-ray band.  For our parameters, this
reduces to B$\delta^{1/3}$=1.1 mG, and for $\delta$=5, 0.6 mG, a value
reasonably consistent with the 1 mG derived on the basis of
equipartition conditions for HST-1 before the major flare (Paper I).


If $E^2$ losses were the controlling factor in the light curve decay,
we would expect $\tau$(UV) to be longer by the square root of the
ratio of the frequencies, $\approx$14.  The actual observed $\tau$(UV)
is $1/0.85=1.18$ yr although the uncertainties include much longer times
($fpy$=0 is within the 1$\sigma$ error).  Of course the fact that both
the UV and the radio intensities decline significantly at later times
indicates that expansion of the source is probably a major contributor
to the light curves behavior during some intervals.

\subsection{Comparison of extreme values of $dI/dt$}

\begin{figure}[tb] 
\includegraphics[angle=0,scale=.40]{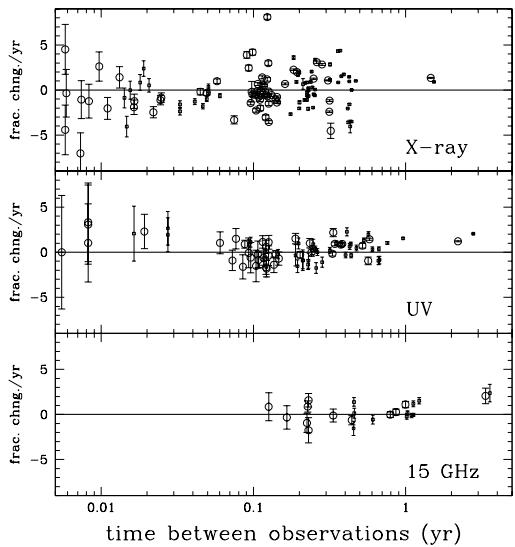}
\caption{The X-ray, UV, and radio scatter plots of $fpy$ values for
HST-1.  The smaller square points are for every other observation and
are thus not independent of the data for adjacent observations (larger
circles).  Top to bottom: X-ray, UV, and 15 GHz.  The absence of large
values at large times in the top panel arises from the smaller
intervals between observations: if we had calculated $fpy$ for every
possible pair of observations, we would have recovered the data
describing the giant flare.}\label{fig:roxfpy}
\end{figure}

In fig.~\ref{fig:roxfpy} we plot the $fpy$ values for HST-1 in different 
bands.  Although many of the uncertainties are large, and the sampling in 
the radio is clearly insufficient, we find that the largest (absolute 
value) believable negative $fpy$'s are --5, --2, and --1 for the X-ray, 
UV, and radio, respectively.  We take this as evidence that energy loss by 
expansion alone, is not indicated.  The obvious caveat to this conclusion 
is that there is a spectral break in the optical/UV in the sense that 
$\alpha_{\rm ox}>\alpha_{\rm ro}$. If there is 
additionally a curving downwards of the spectrum in the $\nu=10^{16}$ to 
10$^{18}$ Hz band, we could explain the $fpy$ data with just expansion.

\section{Discovery of Impulsive Brightening in HST-1 via 
$dI/dt$}\label{sec:impulse}

\begin{figure}[tb] 
\includegraphics[angle=0,scale=.40]{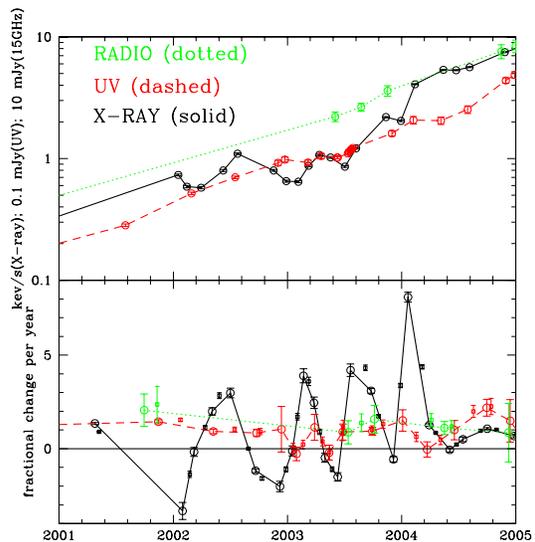}
\caption{Upper panel: lightcurves of HST-1: X-ray (solid), UV (dashed)
and radio (dotted).  Lower panel: $fpy$ values for HST-1 prior to the
giant flare.  Larger circles are the primary data from adjacent
observations.  Smaller circles are generated from every other
observation.  The curves connect primary data.}
\label{fig:rox2001fpy}

\end{figure}

Close examination of figs.~\ref{fig:4fpy} and \ref{fig:rox2001fpy}
(upper panel) shows that the X-ray light curve has a series of small
peaks superposed on a gradually rising intensity between 2002 and 2004.
Their reality is demonstrated by the plot of the first derivative
(figs.~\ref{fig:4fpy} \& \ref{fig:rox2001fpy}).  Characteristic times
of this oscillation (peak-to-peak or trough-to-trough) range from 0.50
years (most common) to a maximum value of 0.84 year.  Although there
are some slightly discernible features on the $fpy$(UV) plots, the
data are not sufficiently numerous or robust enough to search for lags
between bands.  The impulses are not evident at radio frequencies and
their existence is debatable for the UV although if they were as
large in the UV as in the X-ray, they should have been detected.

Although the causes of these oscillations are not known, we speculate
that a quasi periodic variation in the conversion of bulk kinetic jet
power to the internal energy of the radiating plasma is more likely
than a modulation of power flowing down the jet.  We also disfavor a
changing beaming factor caused by a changing angle to the line of
sight (e.g. a ``thrashing jet'').  For jet modulation, a thrashing
jet, and periodic compression and expansion, we would expect any
oscillation to be evident at all frequencies equally.  The absence of
oscillations in the UV encourages us to look for an oscillating
injection of particles which, for the highest energy electrons is made
manifest by the short lifetimes, but for the electrons radiating at
lower frequencies, gets smoothed out by the continuously accumulating
total number of radiating particles.

\section{Comparing X-ray timescales of HST-1 and the nucleus to evaluate 
the site of TeV flaring}\label{sec:compare}

To date, there have been two reports of TeV short-timescale flaring
from M87: 2005 Apr \citep[HESS,][]{ahar06} and 2008 Feb
\citep[MAGIC,][]{albe08}; \citep[VERITAS, e.g.][] {veritas08}.  While the
angular resolution of the TeV systems does not permit a location to be
determined, the expected relation between X-ray intensity and TeV
intensity via an inverse Compton model holds the potential of
localizing the site of the TeV emission if an unambiguous feature in
the TeV light curve can be associated with one in the X-rays.
Moreover, since both instances of TeV flaring appeared to be
characterized by timescales of only a few days, it is also possible to
evaluate statistical differences in X-ray timescales, particularly for
the two leading contenders, the nucleus and HST-1.

\begin{figure}[tb] 
\includegraphics[angle=0,scale=.40]{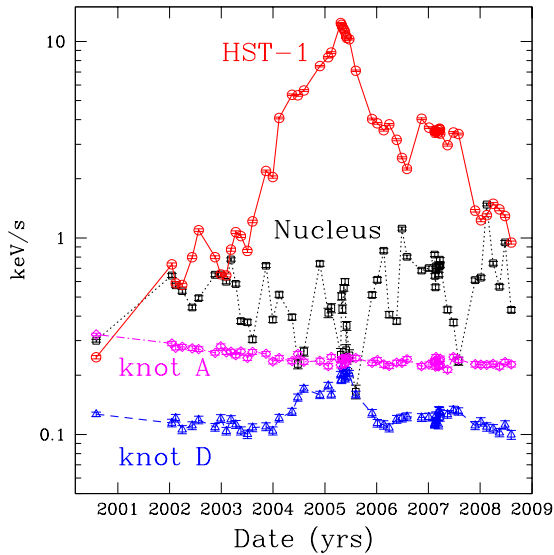}
\caption{X-ray lightcurves for the nucleus (squares), HST-1 (circles),
knots D (triangles) and A (diamonds).  5\% of the intensity of HST-1
has been subtracted from the nuclear values.  We consider it highly
probable that most/all of the knot D apparent variability is simply
contamination from HST-1, and the slight shift to later time of the
peak in 2005 is caused by the secondary response (release of trapped
charge) of the HST-1 PSF.  The secular decline of the knot A
lightcurve is roughly consistent with the loss of effective area at
low energies caused by contamination buildup on the ACIS filter.
\label{fig:4lc}}

\end{figure}

There are a few striking differences in fig.~\ref{fig:4fpy} between
the nucleus and HST-1.  For the nucleus, there are quite large values
of $fpy$ at short time scales, whereas HST-1 has an order of magnitude
smaller amplitudes and these occur at somewhat longer timescales.  We
interpret the presence of large amplitudes at short sampling times
together with smaller amplitudes at longer times to mean that we can
characterize the nuclear variability as a sort of 'flickering'.  HST-1
of course provided us with a major flare with a timescale of a year or
more (figs.~\ref{fig:roxlc},\ref{fig:4lc}).  The energy emitted by the
nuclear flickering is a small fraction of the energy emitted by the
flaring of HST-1, but the timescales are quite different.

\begin{figure}[tb] 
\includegraphics[angle=0,scale=.40]{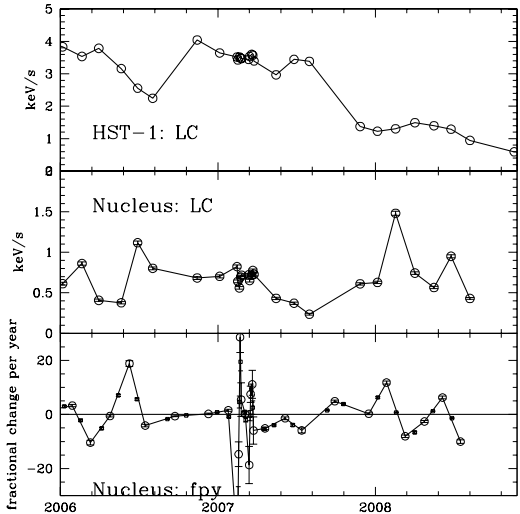}
\caption{The first derivative of the nucleus from 2006 during which
time HST-1 was $<$4 keV/s.  The top panel is a segment of the light
curve of HST-1 and the middle panel shows the light curve for the
nucleus with 5\% of HST-1 subtracted.  There are no egregious
signatures of contamination.  The bottom panel shows the first
derivative of the nuclear lightcurve with smaller squares for
intensity pairs with an intervening
observation.}\label{fig:cchmed3pan}
\end{figure}

In Paper~IV, we argued that the site of the TeV flaring observed by
HESS could be HST-1, whereas others [e.g. \citet{geor05},
  \citet{tav08}, and see also \citet{lev00}] have
suggested a location closer to the super massive black hole (SMBH).
We pointed out that besides the coincidence of the peak of the HST-1
(X-ray, UV, radio) light curve occurring at the same time as the HESS
event in 2005, we knew that HST-1 was physically small (from the
\VLBA\ observations and from the X-ray variability), that the emitted
power at TeV energies was comparable to the X-ray power (linked,
for example, by an SSC model), and that there were difficulties of
getting TeV photons out from the immediate vicinity of the SMBH.
While none of these considerations have changed, our current analyses
on timescales could be thought of as circumstantial evidence that the
nuclear X-ray emission comes from a smaller emitting volume than that
of HST-1 and that this smaller emission region could be the same as
that producing the TeV flares, even though our upper limits on the
size of the x-ray emitting region are still much larger than the light
day timescales inferred for the TeV region.

Early in 2008 Feb we detected an increase in the nuclear X-ray
emission to a level a bit higher than it has ever been
(figs.~\ref{fig:cchmed3pan},\ref{fig:4lc}). This single \Chandra\
observation was made during the TeV flaring observed by MAGIC and
VERITAS and the corresponding values of $fpy$ were \fpyp=+11.8$\pm$0.6 and
\fpym=-8.0$\pm$0.5.  So far in 2008, HST-1 has been at a low level with only
small changes in amplitude.  Unfortunately we do not have a good
estimate of X-ray timescales for the 2005 HESS event.  Although we had
a series of weekly observations, the large fyp(nuclear) values could
have been contaminated by rapid changes in HST-1, as indicated by
\fpy(HST-1) measurements derived from the readout streak photometry.
If there is substantial TeV flaring in 2009, we may be able to settle
this question via an approved target of opportunity Chandra proposal
which aims to find correlated X-ray/TeV behavior in the respective
light curves.

\section{Summary}

We have found a quasi-periodic impulsive signature in the brightening
and dimming of HST-1 in the X-rays.  While this could be interpreted
as a manifestation of past modulation of jet power, we suspect that it
is rather a local oscillation of the process that converts bulk
kinetic jet power to the internal energy of the emitting plasma.  The
fact that HST-1 lies on the northern edge of the cone defined by the
\VLBA\ jet (Paper IV) and that the cross section of HST-1 is less than
0.1\% of the jet area (the cone of the VLBA jet has a diameter of
$\approx$ 90 mas at the distance of HST-1 whereas the effective
diameter of the unresolved upstream end of HST-1 is $\approx$ 2.5
mas), leads us to speculate that the time varying acceleration of
electrons is related to a local instability.  The ratio of $<$0.1\% in
areas is consistent with the ratio of power emitted by HST-1 to that
believed to be the kinetic power of the jet \citep{bick96}.

The finding that the decay times of the lightcurves of HST-1
progressively lengthen moving from high to low frequencies suggests
that simple expansion is not the primary energy loss mechanism for the
relativistic electrons.  If the X-ray decay time actually reflects the
synchrotron halflife of the highest energy electrons, we are not only
able to estimate a value of the magnetic field which is independent of
the usual equipartition assumptions, but also to provide an
explanation of why the impulsive brightening is seen only at X-rays.
That is because the period of the oscillations is very similar to the
X-ray decay time.  The UV decay time is of order 10 times longer, so
that an oscillating brightening would be smoothed out by the failure
of the UV radiating electrons to lose their energy before the next
brightening.  The similarity of the decay time to the oscillatory
brightening (both at X-ray frequencies) may be a coincidence (which
makes the effect manifest), or it could be a causal component of the
(as yet to be determined) instability mechanism.

The X-ray variability timescale evidence suggests that the nucleus
displays faster variability than does HST-1. This is circumstantial
evidence for the hypothesis that the site of the flaring TeV emissions
is the unresolved nucleus rather than HST-1. However, the shortest
nuclear variability timescale we can measure from the {\it Chandra}
data ($\leq 20$\,days) is still significantly longer than the shortest
TeV variability of M~87 reported by the HESS and MAGIC telescopes ($
1-2$\,days).

The analogy we find useful in thinking about HST-1 is that of a river
representing the underlying power flow of the jet.  When something
occurs to transfer some fraction of this power flow into a radiating
plasma, we think of this as 'white water'.  While knot A might be
compared to a waterfall, HST-1 is more like a rock in the river.  Thus
the giant flare could have been caused by a change in local conditions
(e.g. something moving into the river), with the resulting bits of
white water carried downstream as the observed radio blobs.

\acknowledgements

We acknowledge the \Chandra\ Director's Office for approving our
request to obtain a series of observations with short intervals in
2007 Feb and Mar.  Those data were of critical importance in
distinguishing timescale differences between the nucleus and HST-1.
A. S. Wilson (deceased) \& W. Sparks were co-I's on the yearly
\Chandra\ proposals.  We have benefited from discussions with
R. Blandford; with colleagues in the 'TeV community', particularly
H. Krawczynski, P.  Collin, and M. Hui; with F. Primini on aspects of
error propagation; and with B. Wargelin, G. Allen, and R. Edgar for
advice on the subtle aspects of the ACIS CCD.  We thank the anonymous
referee for a useful critique.

Work at SAO was supported by NASA grants GO6-7112X, GO7-8119X, and 
GO8-9116X.  C.C.C. acknowledges support from NRAO through a Jansky 
Postdoctoral Fellowship (2004-2007) and the NASA Postdoctoral Program at 
Goddard Space Flight Center, administered by Oak Ridge Associated 
Universities through a contract with NASA. \L.~S. acknowledges support 
by MEiN through the research project 1-P03D-003-29 in years 2005-2008.
E.S.P. acknowledges support from the NASA LTSA program through grant 
NNX07AM17G.

{\it Facilities:} \facility{VLA}, \facility{HST}, \facility{CXO (ACIS)}

\appendix

\section{Pileup Problems}\label{sec:pileupprob}

\subsection{Saturation\label{sec:saturation}}

Although the methods of \S~\ref{sec:xray} recover most of the
intensity of piled sources, there are on-board filters that keep
events with energies $>$ 17 keV or with certain bad grades from being
telemetered to the ground.  We suspect that these filters produced a
significant decrease in our measurements when HST-1 was near its peak
intensity, but we have not been able to verify this directly since the
housekeeping files that provide the number of events 'dropped' because
of amplitude and grade appear to be dominated by cosmic ray events.

\subsection{``Eat Thy Neighbor''}\label{sec:etn}

When dealing with emission regions closer than an arcsec to each
other, another pernicious effect of pileup comes from the detection
algorithm.  Whenever a candidate event is found, any other event
within its 3x3 pixel grid (for FAINT MODE) is considered to be a part
of that event.  Of the 2 (or more), the event with the most energy
'wins' and its position determines the reported location, and its
energy is found from the sum of the charges within the 3x3 grid.
HST-1 and the nucleus are separated by 0.86$^{\prime\prime}$ or 1.75
pixels.  Thus in the normal course of events, there will be occasional
conflicts of this sort involving a nuclear count arriving within the
same frame time as one from HST-1.  As the intensity of HST-1 rises,
this will happen more often, and many times the nucleus will 'win'
because it has a harder spectrum than does HST-1.  However, when
pileup gets stronger, HST-1 will 'win' most of the time since almost
all events in a frame time will consist of at least two photons,
boosting the recorded charge so as to exceed the single photon from
the nucleus.  We are unaware of any quantitative estimates of this
effect, but suspect it is not causing any serious problems for our
analyses when HST-1 is $<$ 4 keV/s.

\subsection{Second Order Effects of Pileup\label{sec:second}}

When there is negligible pileup, the mutual overlap of the core and
HST-1 PSFs is $\approx$~5$\pm$2\% for the rectangular regions used for
photometry (fig.~\ref{fig:regions}).  However, when pileup is
significant, there are second order effects which seriously distort
the PSF.  Although we avoid some of these by summing energy instead of
just counts, there are others which cannot be accommodated: e.g., the
release of trapped charge during readout.  Since the most obvious of
these produces a secondary response displaced a few pixels from the
PSF in the direction away from the readout buffer, the primary effect
on adjacent jet features changes with the roll angle.  Given the
celestial position of M87, the roll angle is such that there are long
periods of relatively constant roll angles at the beginning and end of
the M87 viewing season (Nov. to Aug.) and a rather rapid change by
close to 180$^{\circ}$ centered around a date late in March.  Since
the PA of the readout streak is similar to that of the jet at the
beginning and end of the season, the primary second order effect
produces contamination of jet features adjacent to HST-1 in the sense
that prior to March the nucleus is badly affected whereas after the
end of March it is knot D which suffers.  This effect is evident in
fig.~\ref{fig:4lc}.

The fact that this sort of second order effect of pileup has not been
calibrated (nor may be susceptible to calibration) means that changes
in the measured intensity of the nucleus (and knot D) may not be
intrinsic during the time that HST-1 was bright.  For that reason, we
have restricted usage of the core data to times when HST-1 was not too
intense.

\section{X-ray Intensities for the Nucleus and HST-1} 

In Paper III, we gave our measured intensities for HST-1 through
2005.  Here we repeat these and also provide uncertainties and the
values for the nucleus (already with 5\% of HST-1 subtracted), both
through the current season which ended in 2008 Aug.  As described in
Paper III, we can estimate the 'fudge factor', $a$, in the
effective area by measuring the flux from fluxmaps when pileup is not
serious.  However, we plan to deal with spectral properties of the
various features in a future paper so prefer to publish here only our
directly determined intensities.

\begin{deluxetable}{clccrlrl}
\tablecaption{\Chandra\ dates and X-ray intensities of the Nucleus and HST-1
\label{tab:obs}}
\tablewidth{0pt}
\tablehead{
\multicolumn{4}{c}{Observational Parameters} & 
\multicolumn{2}{c}{Nucleus} &
\multicolumn{2}{c}{HST-1} \\

\colhead{Epoch} & \colhead{Date} & \colhead{ObsID} & \colhead{Livetime} & 
\colhead{$K$\tablenotemark{a}} & \colhead{$\sigma$\tablenotemark{b}}  & \colhead{$K$\tablenotemark{a}} & \colhead{$\sigma$\tablenotemark{b}}\\
\colhead{label} &  &  & \colhead{(sec)} & \colhead{(keV/s)} &  \colhead{(keV/s)} &  \colhead{(keV/s)} &  \colhead{(keV/s)}
}
\startdata
A\tablenotemark{c} & 2000 Jul 30  & 1808  & 12845    & 0.300   &  0.005    &   0.247     & 0.005  \\ 
B   & 2002 Jan 16  & 3085  & 4889   & 0.644   &  0.014    &   0.734     & 0.013  \\
C   & 2002 Feb 12  & 3084  & 4655   & 0.575   &  0.013    &   0.588     & 0.012  \\
D   & 2002 Mar 30  & 3086  & 5089   & 0.536   &  0.013    &   0.576     & 0.012  \\
E   & 2002 Jun 08  & 3087  & 4973   & 0.443   &  0.012    &   0.798     & 0.014  \\
F   & 2002 Jul 24  & 3088  & 4708   & 0.494   &  0.013    &   1.096     & 0.017  \\
G   & 2002 Nov 17  & 3975  & 5287   & 0.649   &  0.014    &   0.799     & 0.013  \\
H   & 2002 Dec 29  & 3976  & 4792   & 0.652   &  0.014    &   0.653     & 0.013  \\
I   & 2003 Feb 04  & 3977  & 5276   & 0.598   &  0.013    &   0.645     & 0.012  \\
J   & 2003 Mar 09  & 3978  & 4852   & 0.777   &  0.016    &   0.872     & 0.015  \\
K   & 2003 Apr 14  & 3979  & 4492   & 0.583   &  0.014    &   1.071     & 0.017  \\
L   & 2003 May 18  & 3980  & 4788   & 0.378   &  0.011    &   1.022     & 0.016  \\
M   & 2003 Jul 03  & 3981  & 4677   & 0.372   &  0.011    &   0.859     & 0.015  \\
N   & 2003 Aug 08  & 3982  & 4841   & 0.304   &  0.010    &   1.214     & 0.018  \\
O   & 2003 Nov 11  & 4917  & 5028   & 0.723   &  0.016    &   2.192     & 0.026  \\
P   & 2003 Dec 29  & 4918  & 4677   & 0.384   &  0.012    &   2.041     & 0.026  \\
Q   & 2004 Feb 12  & 4919  & 4703   & 0.515   &  0.016    &   4.079     & 0.04   \\
R\tablenotemark{d} & 2004 Mar 29 & 4920 & 5235 & \nodata & \nodata & \nodata & \nodata  \\
S   & 2004 May 13  & 4921  & 5251   & 0.396   &  0.013    &   5.358     & 0.05   \\
T   & 2004 Jun 23  & 4922  & 4543   & 0.228   &  0.013    &   5.323     & 0.05   \\
U   & 2004 Aug 05  & 4923  & 4633   & 0.264   &  0.014    &   5.636     & 0.055  \\
V   & 2004 Nov 26  & 5737  & 4237   & 0.740   &  0.022    &   7.494     & 0.07   \\
W   & 2005 Jan 24  & 5738  & 4666   & 0.416   &  0.023    &   8.316     & 0.08   \\
X   & 2005 Feb 14  & 5739  & 5154   & 0.439   &  0.024    &   8.785     & 0.08   \\
Ya  & 2005 Apr 22  & 5740  & 4699   & 0.254   &  0.019    &   12.417    & 0.11   \\
Yb  & 2005 Apr 28  & 5744  & 4699   & 0.505   &  0.022    &   12.167    & 0.11   \\
Yc  & 2005 May 04  & 5745  & 4705   & 0.435   &  0.021    &   11.798    & 0.108  \\
Yd  & 2005 May 13  & 5746  & 5142   & 0.557   &  0.022    &   11.555    & 0.10   \\
Ye  & 2005 May 22  & 5747  & 4701   & 0.597   &  0.023    &   11.243    & 0.10   \\
Yf  & 2005 May 30  & 5748  & 4699   & 0.268   &  0.018    &   10.665    & 0.10   \\
Yg  & 2005 Jun 03  & 5741  & 4698   & 0.357   &  0.019    &   10.432    & 0.095  \\
Yh  & 2005 Jun 21  & 5742  & 4703   & 0.252   &  0.018    &   10.270    & 0.094  \\
Yi  & 2005 Aug 06  & 5743  & 4672   & 0.165   &  0.013    &   7.098     & 0.067  \\
Yj  & 2005 Nov 29  & 6299  & 4655   & 0.514   &  0.016    &   4.032     & 0.042  \\
Yk  & 2006 Jan 04  & 6300  & 4660   & 0.612   &  0.017    &   3.833     & 0.040  \\
Yl  & 2006 Feb 19  & 6301  & 4337   & 0.861   &  0.020    &   3.535     & 0.040  \\
Ym  & 2006 Mar 30  & 6302  & 4701   & 0.407   &  0.014    &   3.786     & 0.040  \\
Yn  & 2006 May 21  & 6303  & 4699   & 0.377   &  0.013    &   3.161     & 0.035  \\
Yo  & 2006 Jun 19  & 6304  & 4677   & 1.117   &  0.021    &   2.557     & 0.031  \\
Yp  & 2006 Aug 02  & 6305  & 4653   & 0.802   &  0.018    &   2.246     & 0.028  \\
Yq  & 2006 Nov 13  & 7348  & 4543   & 0.682   &  0.018    &   4.042     & 0.04   \\
Yr  & 2007 Jan 04  & 7349  & 4685   & 0.703   &  0.018    &   3.642     & 0.039  \\
Ys  & 2007 Feb 13  & 7350  & 4662   & 0.823   &  0.019    &   3.516     & 0.039  \\
Yt  & 2007 Feb 15  & 8510  & 4701   & 0.641   &  0.017    &   3.428     & 0.038 \\
Yu  & 2007 Feb 18  & 8511  & 4703   & 0.561   &  0.016    &   3.515     & 0.038  \\
Yv  & 2007 Feb 21  & 8512  & 4703   & 0.694   &  0.017    &   3.479     & 0.038 \\
Yw  & 2007 Feb 24  & 8513  & 4700   & 0.717   &  0.018    &   3.472     & 0.038 \\
Yx  & 2007 Mar 12  & 8514  & 4471   & 0.722   &  0.018    &   3.443     & 0.039  \\
Yy  & 2007 Mar 14  & 8515  & 4696   & 0.651   &  0.017    &   3.533     & 0.038  \\
Yz  & 2007 Mar 19  & 8516  & 4679   & 0.716   &  0.018    &   3.599     & 0.039 \\
Za  & 2007 Mar 22  & 8517  & 4674   & 0.775   &  0.019    &   3.571     & 0.039 \\ 				    
Zb  & 2007 Mar 24  & 7351  & 4683   & 0.728   &  0.018    &   3.396     & 0.039 \\
Zc  & 2007 May 15  & 7352  & 4588   & 0.432   &  0.014    &   2.968     & 0.034 \\
Zd  & 2007 Jun 25  & 7353  & 4543   & 0.372   &  0.014    &   3.445     & 0.038  \\
Ze  & 2007 Jul 31  & 7354  & 4707   & 0.236   &  0.011    &   3.383     & 0.037 \\
Zf  & 2007 Nov 25  & 8575  & 4679   & 0.610   &  0.015    &   1.376     & 0.020  \\
Zg  & 2008 Jan 05  & 8576  & 4692   & 0.628   &  0.015    &   1.230     & 0.019  \\
Zh  & 2008 Feb 16  & 8577  & 4659   & 1.480   &  0.024    &   1.305     & 0.020 \\
Zi  & 2008 Apr 01  & 8578  & 4706   & 0.743   &  0.017    &   1.493     & 0.021 \\
Zj  & 2008 May 15  & 8579  & 4706   & 0.565   &  0.014    &   1.398     & 0.021 \\
Zk  & 2008 Jun 24  & 8580  & 4705   & 0.950   &  0.019    &   1.293     & 0.020 \\
Zl  & 2008 Aug 07  & 8581  & 4657   & 0.431   &  0.012	  &   0.947     & 0.016  \\
			
\enddata			
				 


\tablenotetext{a}{The values of $K$ are 'detector-based' observed intensities.
  They come from summing the energies of all events (from the 'evt1'
  file) within a rectangle of length 5 pixels transverse to the jet
  and 2.5 pixels along the jet (fig.~\ref{fig:regions}).  
  No background subtraction was employed: when HST-1 is weak, the
  background (in counts) is of order 1\%; for the highest
  intensity, the background is less than 0.3\%.  There is no
  correction for the build-up of ACIS contamination.}

\tablenotetext{b}{Uncertainties are based on the raw counts: $\sqrt{N}/N$, hence 1$\sigma$.}

\tablenotetext{c}{Archival data from \citet{wils02}.}

\tablenotetext{d}{This observation was taken in continuous clocking
  mode, so there is no 2D image available. All other observations had 1/8th subarray ACIS-S3 chip only,
  and 0.4s frame time.}

\end{deluxetable}


{}

\end{document}